\documentclass[a4paper,11pt]{article}

\usepackage{pos}
\usepackage{amsmath}
\usepackage{dsfont}
\usepackage{slashed}
\usepackage[utf8]{inputenc}
\usepackage{graphicx}
\usepackage{subcaption}
\usepackage{hyperref}

\graphicspath{{./figures/}}

\newcommand{\Ns}{N_{\mathrm{s}}}
\newcommand{\Nt}{N_{\mathrm{t}}}
\newcommand{\Nf}{N_{\mathrm{f}}}
\newcommand{\tr}{\mathrm{tr}}
\newcommand{\ii}{\mathrm{i}}
\newcommand{\g}{\gamma_5}
\newcommand{\cc}{\langle\bar{\psi}\psi\rangle}
\newcommand{\sigex}{\langle\sigma\rangle}
\newcommand{\Dov}{D_\mathrm{ov}}
\renewcommand{\L}{\mathcal{L}}
\newcommand{\id}{\mathds{1}}

\title{Magnetic catalysis in the $1$-flavor Gross-Neveu model in $2+1$ dimensions}
\ShortTitle{Magnetic catalysis in the 3D GN model}

\author*[a]{Michael Mandl}
\author[a,b]{Julian J. Lenz}
\author[a]{Andreas Wipf}

\affiliation[a]{Theoretisch-Physikalisches Institut, Friedrich-Schiller-Universität Jena,\\
Fröbelstieg 1, D-07743 Jena, Germany}
\affiliation[b]{Swansea Academy of Advanced Computing, Swansea University,\\
Fabian Way, SA1 8EN, Swansea, Wales, UK}

\emailAdd{j.j.lenz@swansea.ac.uk}
\emailAdd{michael.mandl@uni-jena.de}
\emailAdd{wipf@tpi.uni-jena.de}

\abstract{We investigate the Gross-Neveu model in $2+1$ dimensions in a constant and homogeneous magnetic field using one reducible flavor of overlap fermions. Our lattice simulations suggest that the magnetic catalysis phenomenon, i.e., an increase of the chiral condensate with the magnetic field, is present for all temperatures below the chiral phase transition, in accordance with predictions from mean-field and beyond-mean-field calculations.}

\FullConference{%
The 39th International Symposium on Lattice Field Theory,\\
8th-13th August, 2022,\\
Rheinische Friedrich-Wilhelms-Universität Bonn, Bonn, Germany
}

\begin{document}
\maketitle

\section{Introduction}
Strong magnetic fields are known or expected to be present in heavy-ion collisions, neutron stars and the early universe \cite{MS15r}. Thus, a proper understanding of their influence on the structure of matter, both at finite temperature and at finite density, is of great relevance. The framework best suitable to describe these phenomena is quantum chromodynamics (QCD), the theory of strong interactions. However, QCD at finite density is plagued by the infamous complex-action problem, preventing the use of conventional Monte-Carlo methods. One way to, at least partially, circumvent this issue is to resort to the use of low-energy effective theories or toy models. 

Examples for such toy models are four-Fermi theories like the Gross-Neveu (GN) model \cite{GN74} and some of its variants, which have not only been used successfully to model low-energy properties of QCD but can also be found in abundance in the context of condensed-matter physics \cite{CM94,EB19}. These models have, so far, most commonly been studied in the mean-field approach, which is equivalent to the limit where the number of fermion flavors tends to infinity and quantum fluctuations are suppressed. 

In this contribution we first summarize the most important features of the $(2+1)$-dimensional GN model at finite temperature, density and magnetic field, obtained in the mean-field limit. We then take a first step towards the goal of studying QCD in a magnetic field via effective models by investigating the full GN model at finite flavor number on the lattice, albeit restricting for now to vanishing chemical potential. The case of finite density will then be studied in a forthcoming publication.

\section{The Gross-Neveu model}
The Gross-Neveu model is defined by the Lagrangian
\begin{align}\label{eq:4F_Lagrangian}
	\L = \bar{\psi}\ii\slashed{\partial}\psi + \frac{g^2}{2\Nf}(\bar{\psi}\psi)^2\;,
\end{align}
where $g^2$ is a coupling constant and $\Nf$ denotes the number of fermionic flavors, which are summed over implicitly in \eqref{eq:4F_Lagrangian}. With the help of an auxiliary scalar field $\sigma$, one can bring \eqref{eq:4F_Lagrangian} into an equivalent form, which reads, after introducing in the usual ways a chemical potential $\mu$ and an external vector field $A_\mu$,
\begin{align}\label{eq:Lagrangian}
	\L = \ii\bar{\psi}(\slashed{\partial}+\sigma+\mu\gamma_0+ie\slashed{A})\psi + \frac{\Nf}{2g^2}\sigma^2\;.
\end{align}
Here, $e$ denotes the elementary electric charge. This model has a discrete chiral symmetry,
\begin{align}\label{eq:continuum_symmetry}
	\psi\rightarrow\ii\g\psi\;, \quad \bar{\psi}\rightarrow\ii\bar{\psi}\g\;, \quad \sigma\rightarrow-\sigma\;,
\end{align}
which may be spontaneously broken by the formation of a chiral condensate $\cc$. The latter can be shown to be proportional to the expectation value of $\sigma$, i.e.
\begin{align}\label{eq:ward_identity}
	\cc = \frac{\ii\Nf}{g^2}\sigex\;.
\end{align}
Apart from spontaneous chiral symmetry breaking, the GN model also shares other important features with QCD, such as renormalizability (in three dimensions or lower) and asymptotic freedom (in two dimensions).

\section{The large-\texorpdfstring{$\Nf$}{Nf} limit}
\begin{figure}[h]
	\begin{subfigure}{0.33\textwidth}
		\centering
		\includegraphics[scale=0.327]{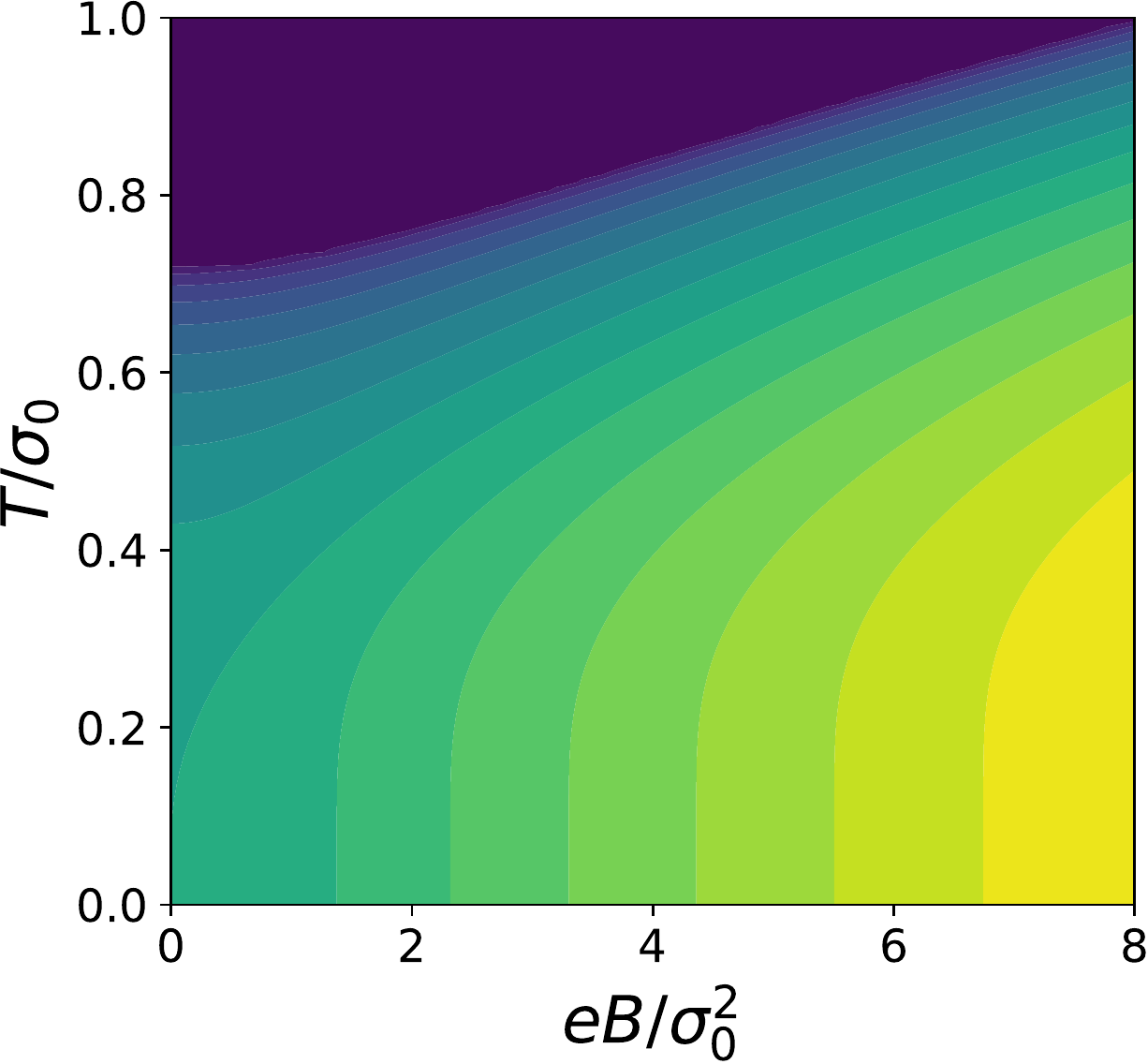}
		\caption{$\mu=0$}
	\end{subfigure}
	\hspace{-0.6cm}
	\begin{subfigure}{0.33\textwidth}
		\centering
		\vspace{0.045cm}
		\includegraphics[scale=0.327]{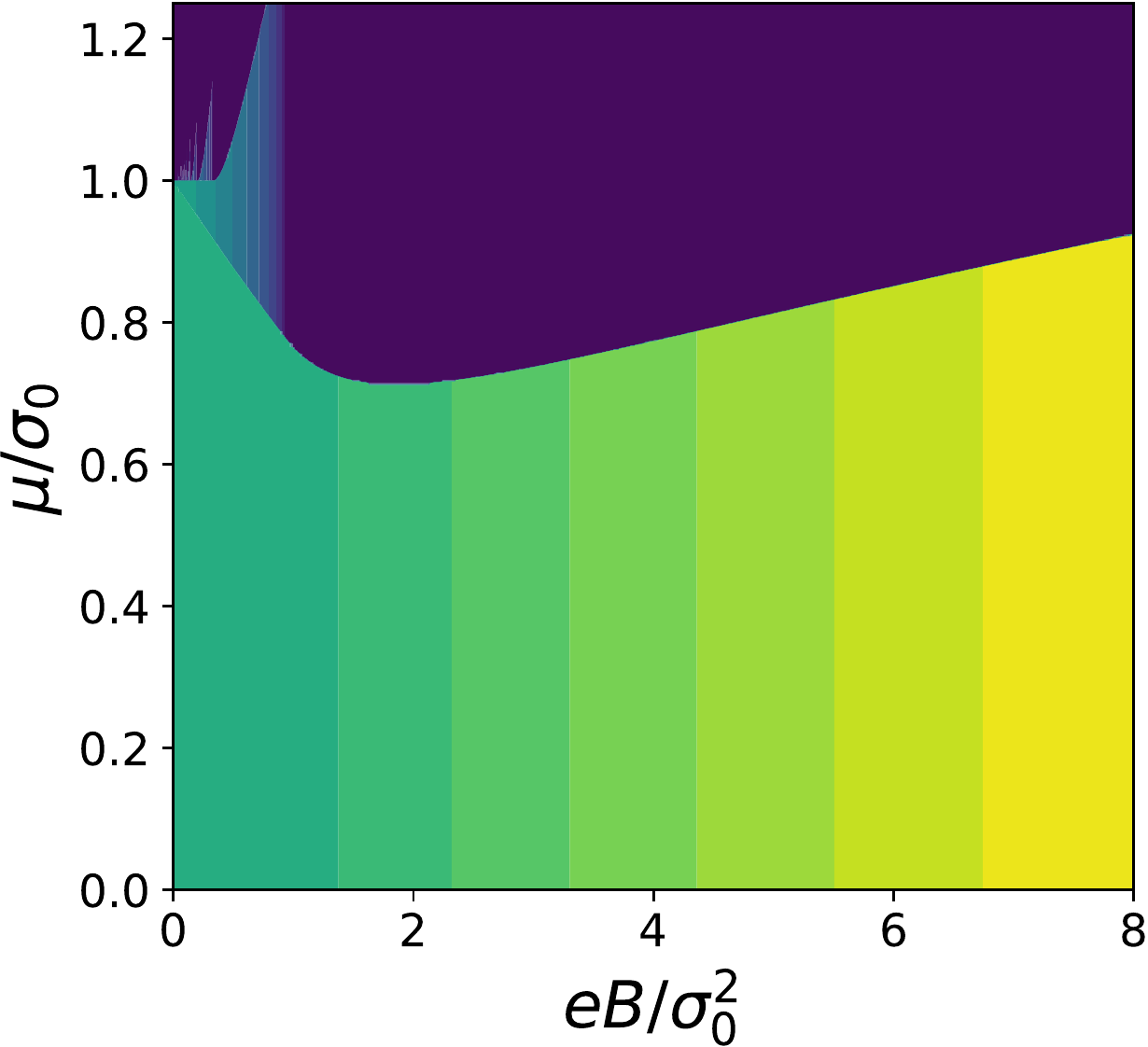}
		\caption{$T=0$}
	\end{subfigure}
	\begin{subfigure}{0.33\textwidth}
		\centering
		\vspace{-0.05cm}
		\includegraphics[scale=0.327]{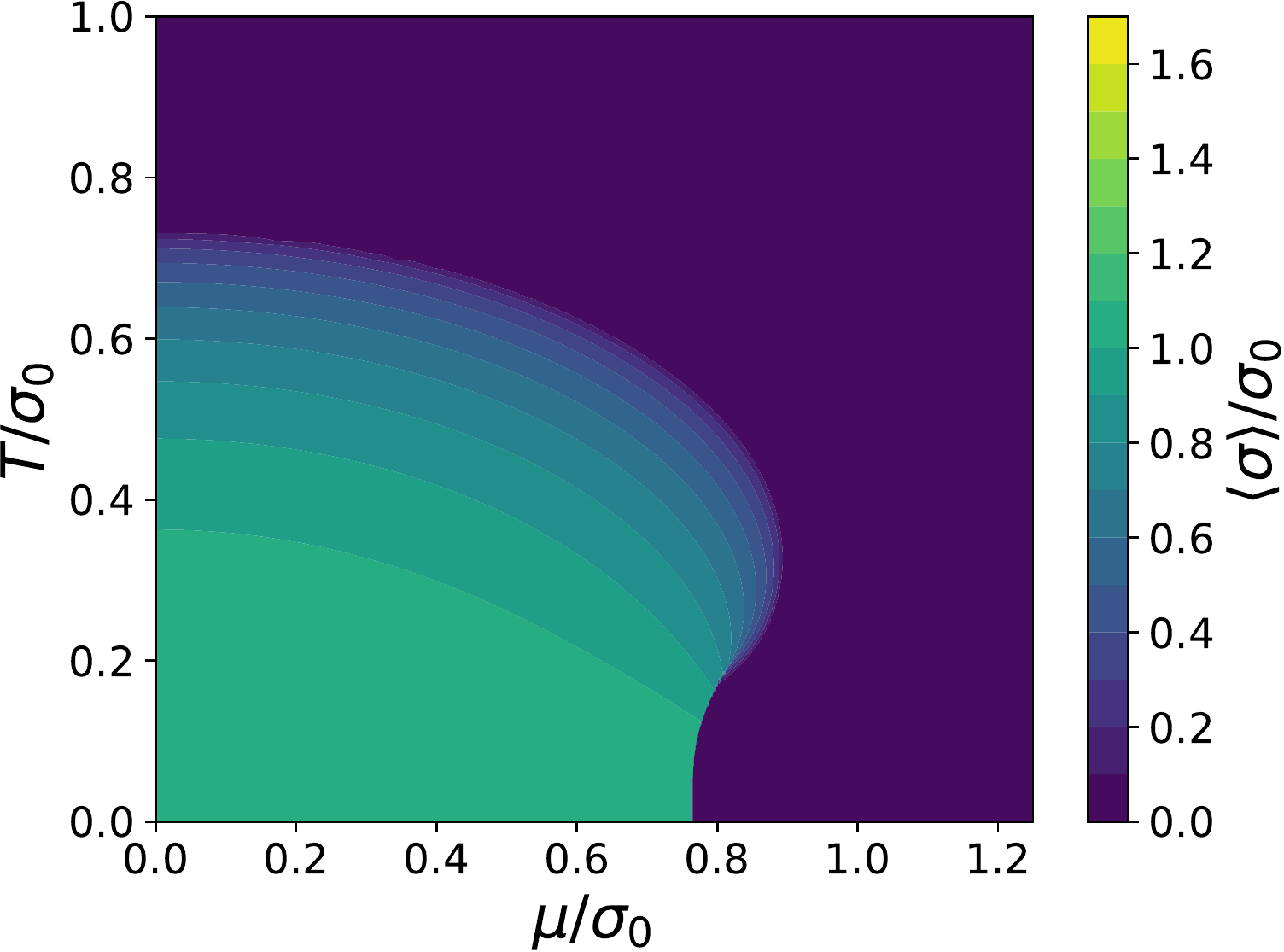}
		\caption{$eB=\sigma_0^2$}
	\end{subfigure}
	\caption{Large-$\Nf$ phase diagrams for various fixed external parameters (see also \cite{KPR13}). The scale $\sigma_0$ denotes the value of the condensate $\sigex$ obtained at vanishing $T$, $\mu$ and $B$.}
	\label{fig:large_Nf}
\end{figure}

In the limit where $\Nf\rightarrow\infty$, computing the path integral in the GN model in $d$ dimensions reduces to a minimization problem of the effective action
\begin{align}\label{eq:action}
	S_{\mathrm{eff}} = \frac{1}{2g^2}\int d^dx\,\sigma^2(x)-\tr\log D[\sigma]\;,
\end{align}
with the Dirac operator
\begin{align}\label{eq:dirac_operator}
	D[\sigma] = \slashed{\partial} + \sigma +\mu\gamma_0+\ii e\slashed{A}\;.
\end{align}
In fact, in this limit the mean-field approach becomes exact. Restricting to a homogeneous auxiliary field, $\sigma(x)=\sigma$, and assuming the (homogeneous) magnetic field $B$ to lie perpendicular to the spatial plane, the effective action in $2+1$ dimensions, in a reducible $4\times4$ representation of $\gamma$ matrices, can be computed in closed form \cite{GMS95}:
\begin{align}\label{eq:effective_action_large_Nf}
	\begin{aligned}
		\frac{S_\mathrm{eff}}{V} 
	 &= \frac{\sigma^2}{2g_R^2} 
	  - \frac{\sqrt{2}(eB)^{3/2}}{\pi}\zeta\left(-\frac{1}{2}, \frac{\sigma^2}{2eB}\right)
	  + \frac{\sigma eB}{2\pi}\\	    	    		       
	 &- \frac{eB}{2\pi\beta}\sum_{l=0}^{\infty}d_l\left[\log\left(1+e^{-\beta\left(\sqrt{\sigma+2eBl}+\mu\right)}\right) + (\mu\rightarrow -\mu)\right]\;,
	\end{aligned}
\end{align}
where $V$ denotes the volume of the system, $g_R^2$ is the renormalized coupling constant, $\zeta$ denotes the Hurwitz zeta function, $\beta$ is the inverse temperature and $l$ enumerates the discrete Landau levels with their degeneracy factor $d_l=2-\delta_{l0}$. We have assumed for $eB$ to be non-negative. The minimization of \eqref{eq:effective_action_large_Nf} with respect to $\sigma$ reveals a rich phase structure in the parameter space spanned by temperature $T$, chemical potential $\mu$, and magnetic field $B$, see Fig.~\ref{fig:large_Nf}. 

\noindent We summarize the most important observations one makes in the large-$\Nf$ limit as follows:
\begin{itemize}
	\item At $\mu=0$ the chiral condensate $\sigex$ grows monotonically as a function of $B$, i.e., one finds magnetic catalysis \cite{Kli92_2}. The critical temperature $T_c$ of the transition between the spontaneously broken phase (where $\sigex\neq0$) and the symmetric phase (where $\sigex=0$) grows with $B$. 
	
	\item At $T=0$ the condensate experiences magnetic catalysis for some $\mu$ and inverse magnetic catalysis, i.e., a decrease with $B$, for others \cite{PRS11}. The critical chemical potential $\mu_c$ is non-monotonic in $B$. For weak magnetic fields there are multiple first-order phase transitions in $\mu$ between the broken and symmetric phases due to the Landau level structure.
	
	\item For magnetic fields $eB\geq\sigma_0^2$ the lowest Landau level ($l=0$ in Eq.~\eqref{eq:effective_action_large_Nf}) dominates and there are no more multiple transitions. Furthermore, as long as $B$ is not too large, one finds the counter-intuitive situation where increasing the temperature can actually cause a transition from the symmetric to the spontaneously broken phase. 
	
	\item For very strong magnetic fields the condensate grows monotonically with $B$ for all temperatures and chemical potentials, while $T_c$ and $\mu_c$ both increase with $B$.
\end{itemize}

\section{Lattice setup}
We now wish to study what remains of the complicated phase structure shown in Fig.~\ref{fig:large_Nf} when going beyond the large-$\Nf$ limit, i.e., when taking into account bosonic fluctuations. In particular, we study the case of one flavor of (reducible, i.e., $4$-component) fermions, $\Nf=1$. Since we are interested in the chiral properties of the theory, we employ the overlap operator \cite{NN93} in our simulations, namely in the form that was advocated by H. Neuberger \cite{Neu98_2}. We introduce the scalar field $\sigma$ in a way that carries over the Dyson-Schwinger equation \eqref{eq:ward_identity} to the discretized theory and we couple the chemical potential in a chirally-symmetric manner, as suggested by Gavai and Sharma \cite{GS12}. The full Dirac operator in our setup then reads
\begin{align}\label{eq:full_overlap}
  D = \Dov + \left(\sigma+\mu\gamma_0\right)\left(\id-\frac{a}{2}\Dov\right)\;,
\end{align}
where $\Dov$ is the usual massless overlap operator \cite{Neu98_2} with a negative Wilson mass parameter $m=-1$ and $a$ denotes the lattice constant. The magnetic field, which we again choose homogeneous and perpendicular to the spatial plane, is contained in $\Dov$ via $U(1)$ gauge links. These contain appropriately chosen boundary terms, which ensure that the system is physically equivalent to one with a constant magnetic flux $\Phi$ through the entire lattice \cite{AW09}. Moreover, denoting the area of the spatial plane by $L^2$, the flux is quantized in terms of an integer $b$ according to
\begin{align}\label{eq:magnetic_field}
  \Phi = BL^2 = \frac{2\pi}{e} b\;,
\end{align}
due to the periodic boundary conditions on the finite lattice, and $b$ is restricted to a finite range, $b\in\left\{0,\,\dots,\,\Ns^2\right\}$, where $\Ns$ denotes number of lattice points in each spatial direction such that $L=a\Ns$.

In the overlap formalism the continuum $\mathbb{Z}_2$ symmetry \eqref{eq:continuum_symmetry} is mapped to the lattice in the following way:
\begin{align}\label{eq:lattice_symmetry}
  \psi\rightarrow\ii\hat{\gamma}_5\psi\;, \quad \bar{\psi}\rightarrow\ii\bar{\psi}\gamma_5\;, \quad \sigma\rightarrow-\sigma\;,
\end{align}
where $\hat{\gamma}_5=\gamma_5(\id-a\Dov)$. As was mentioned above, one can furthermore show that the identity \eqref{eq:ward_identity} has an exact counterpart in our lattice formulation as well, relating $\sigex$ to the definition of the chiral condensate for Ginsparg-Wilson fermions \cite{Cha99}:
\begin{align}
  \sigex = -\frac{\Nf}{g^2}\left\langle\bar{\psi}\left(\id-\frac{a}{2}\Dov\right)\psi\right\rangle\;.
\end{align}

In the present setup the only dynamical degree of freedom is $\sigma$, which does not enter the definition of the massless overlap operator $\Dov$, involving an expensive operator sign function. We can thus compute this sign function once and for all for a given magnetic field, which reduces the computational cost drastically. In fact, this even allows us to compute $\Dov$ and, by extension, the full operator $D$ exactly, i.e., without relying on any approximations. This would not be feasible in state-of-the-art simulations of gauge theories, where the dynamical degrees of freedom directly enter the sign function, which one then has to re-compute for every update. For a more detailed discussion about our lattice setup we refer to an upcoming publication, in which we also plan to publish our simulation data.

In the following we consider lattices with spatial volumes $\Ns^2\in\{8^2, 12^2,
16^2\}$ and various temporal extents $\Nt=1/aT$ in order to study the
temperature-dependence of the chiral condensate. To avoid cancellations arising
due to the two equivalent minima of the GN effective action present in the phase
of spontaneously broken chiral symmetry, we measure the absolute value of the
auxiliary field, $\langle\vert\sigma\vert\rangle$, as an order parameter.

We set the scale via the value of the order parameter at vanishing $B$ and $\mu$
and at the lowest temperature considered,
\begin{align}\label{eq:lattice_scale}
  \sigma_0=\langle\vert\sigma\vert\rangle_{B=0,\, \mu=0,\, T\approx0}\;.
\end{align}
Since we cannot reach $T=0$ exactly in our simulations, we should, for reasons
of consistency, at least attempt to keep the temperature $T_0\approx0$, at which
we set the scale, constant when approaching the two limits of vanishing lattice
spacing and infinite volume, respectively. We do so in this work, but choose
different values of $T_0$ for the two respective limits in our setup.

We change the lattice spacing $a$ by varying the coupling constant $g^2$,
keeping the physical value of $\sigma_0$ constant in the process, in order to
extrapolate to the continuum limit. We furthermore approach the infinite-volume
limit by increasing the number of spatial lattice points while keeping $a$ and
$\sigma_0$ fixed.

Notice that the theory does not suffer from a complex-action problem at $\mu\neq0$ as long as $B=0$ (and vice versa), which one can prove by using charge-conjugation symmetry. In the combined case, where both the magnetic field and the chemical potential are non-vanishing, however, this argument does not hold anymore and indeed a complex-action problem is present. This case will be discussed in detail elsewhere.

\section{Results}
\begin{figure}[t]
	\begin{subfigure}{0.49\textwidth}
		\centering
		\includegraphics[scale=0.395]{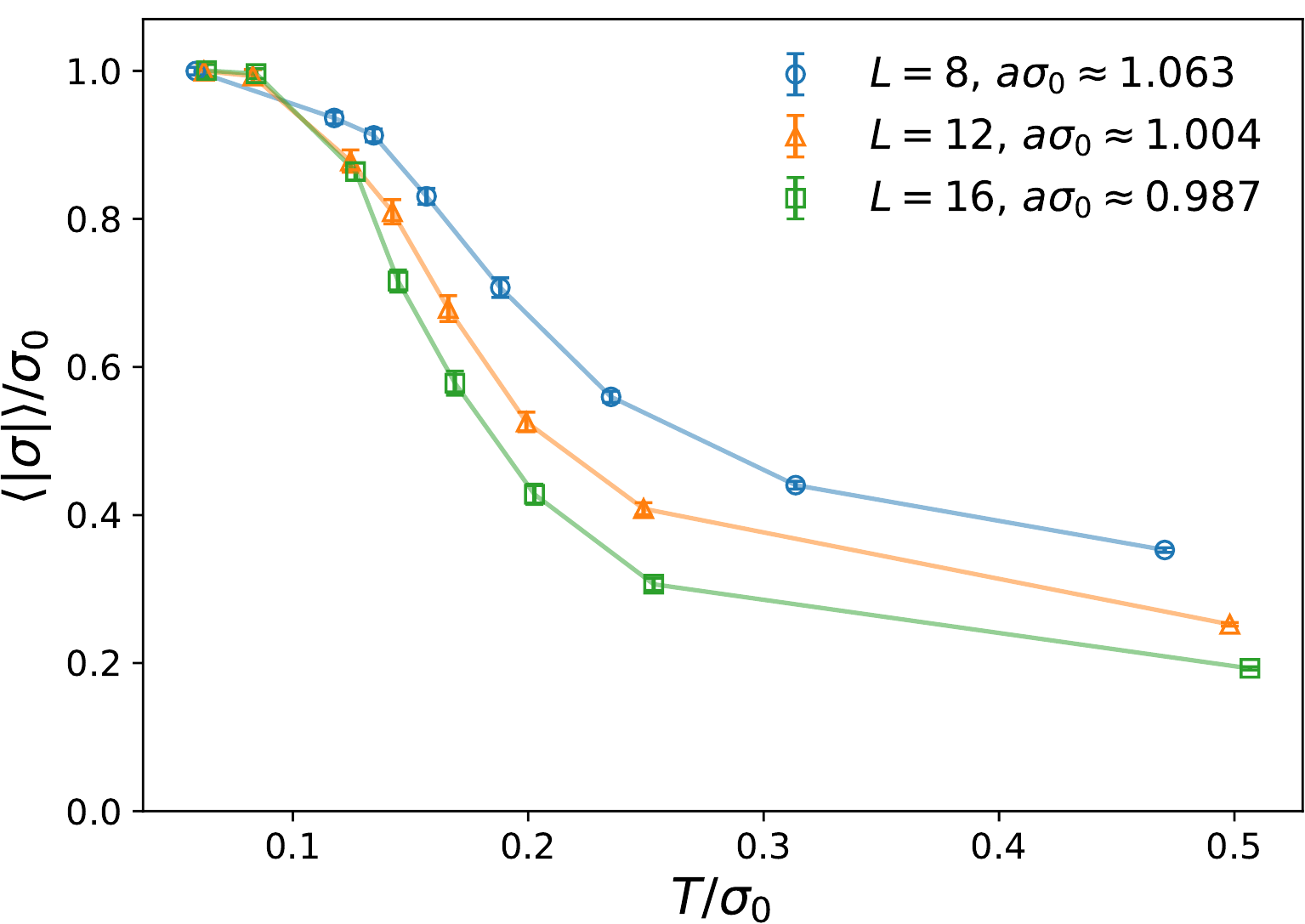}
		\caption{$\mu=0$}
		\label{fig:cc_vs_T_infVol}	
	\end{subfigure}
	\begin{subfigure}{0.49\textwidth}
		\centering
		\includegraphics[scale=0.395]{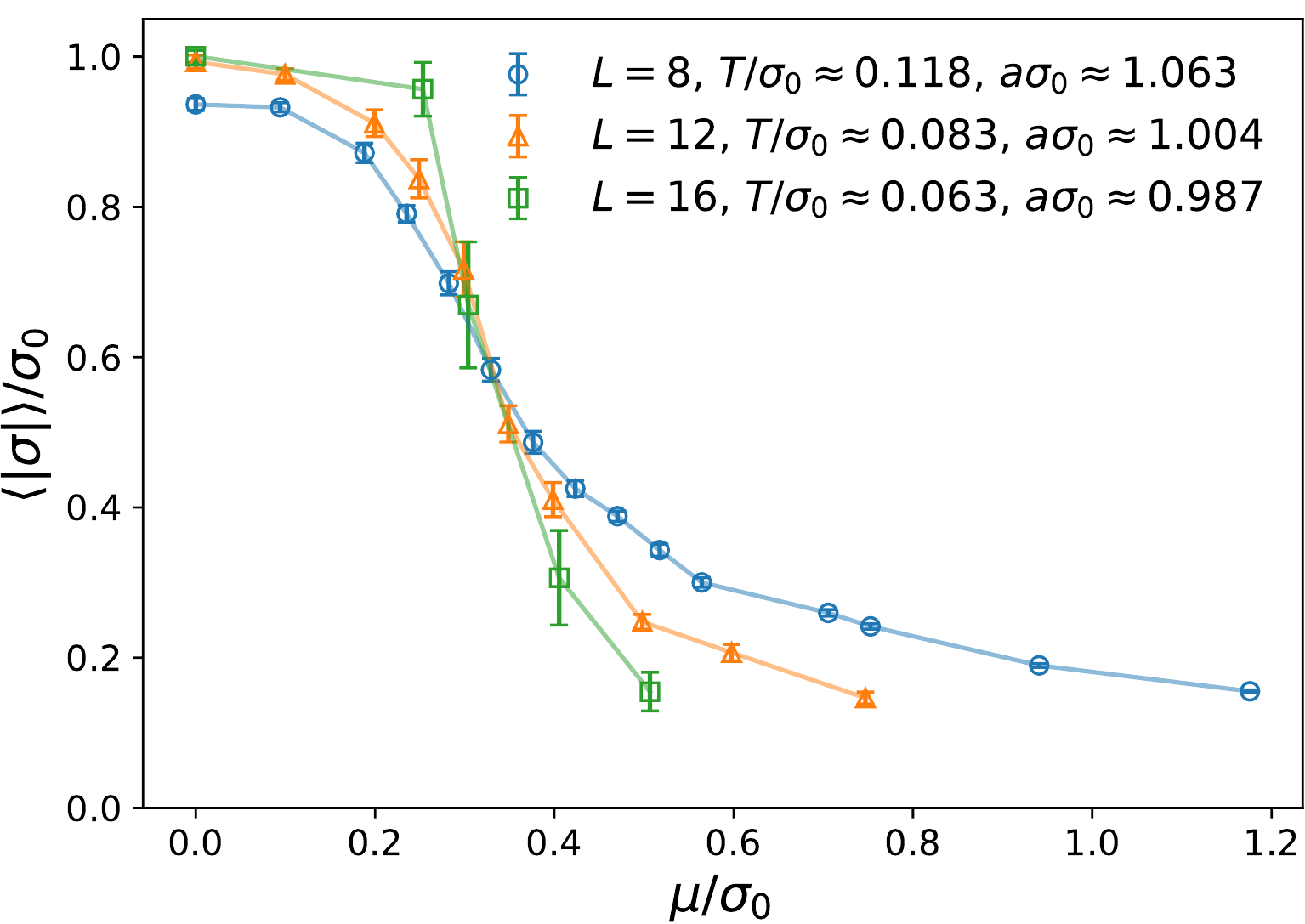}		
		\caption{$T\approx 0$}
		\label{fig:cc_vs_mu_infVol}
	\end{subfigure}
	\caption{Infinite-volume extrapolation of the $T$- and $\mu$-dependence of the order parameter at vanishing magnetic field.}
\end{figure}

As a crosscheck to test the discretization \eqref{eq:full_overlap} we first restrict to the case of vanishing magnetic field, where the model is better understood. In particular, we show the temperature-dependence of the order parameter obtained at $\mu=0=B$ in Fig.~\ref{fig:cc_vs_T_infVol}. One observes the expected behavior of the transition becoming more pronounced as the volume increases, approaching a second-order phase transition for infinitely large volumes. Notice that for finite volume $\langle\vert\sigma\vert\rangle$ will never vanish exactly. 

A similar conclusion can be drawn from the $\mu$-dependence of $\langle\vert\sigma\vert\rangle$ at low temperatures shown in Fig.~\ref{fig:cc_vs_mu_infVol}. Here, mean-field calculations suggest that the transition is of second order for all finite temperatures and of first order (in the Ehrenfest sense) at $T=0$. However, previous lattice studies found that the transition for finite flavor numbers might instead be of weak first order even for non-vanishing temperatures \cite{KS01}, which has been supported by analytical calculations \cite{KPR07}. The small lattice sizes considered in this work obviously do not allow us to determine the order of the phase transition definitively, but the transition should sharpen with increasing volume either way and we indeed observe this.

\begin{figure}[h]
	\begin{subfigure}{0.49\textwidth}
		\centering
		\includegraphics[scale=0.4]{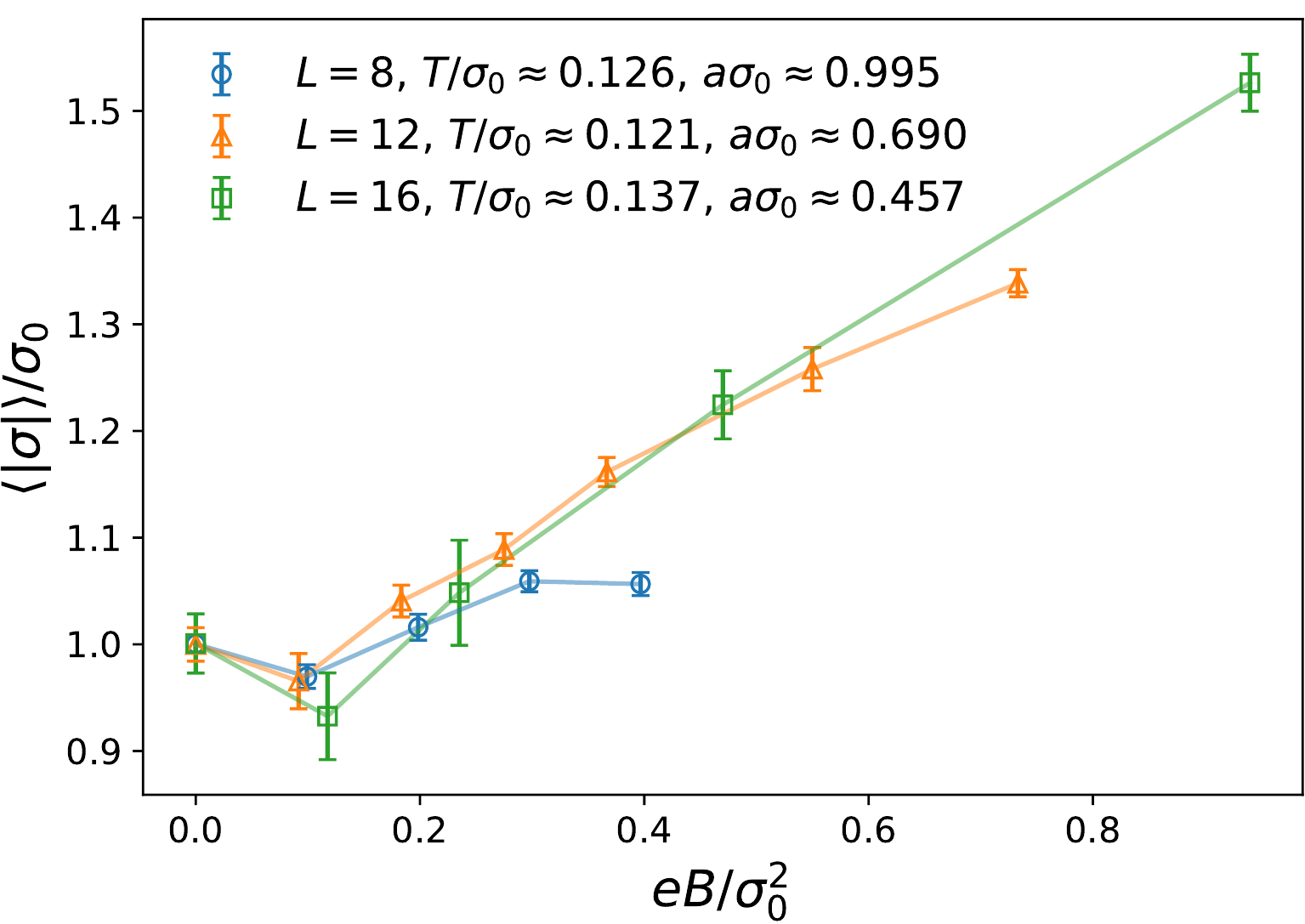}		
		\caption{Continuum extrapolation.}
		\label{fig:cc_vs_B_cont}
	\end{subfigure}
	\begin{subfigure}{0.49\textwidth}
		\centering
		\includegraphics[scale=0.4]{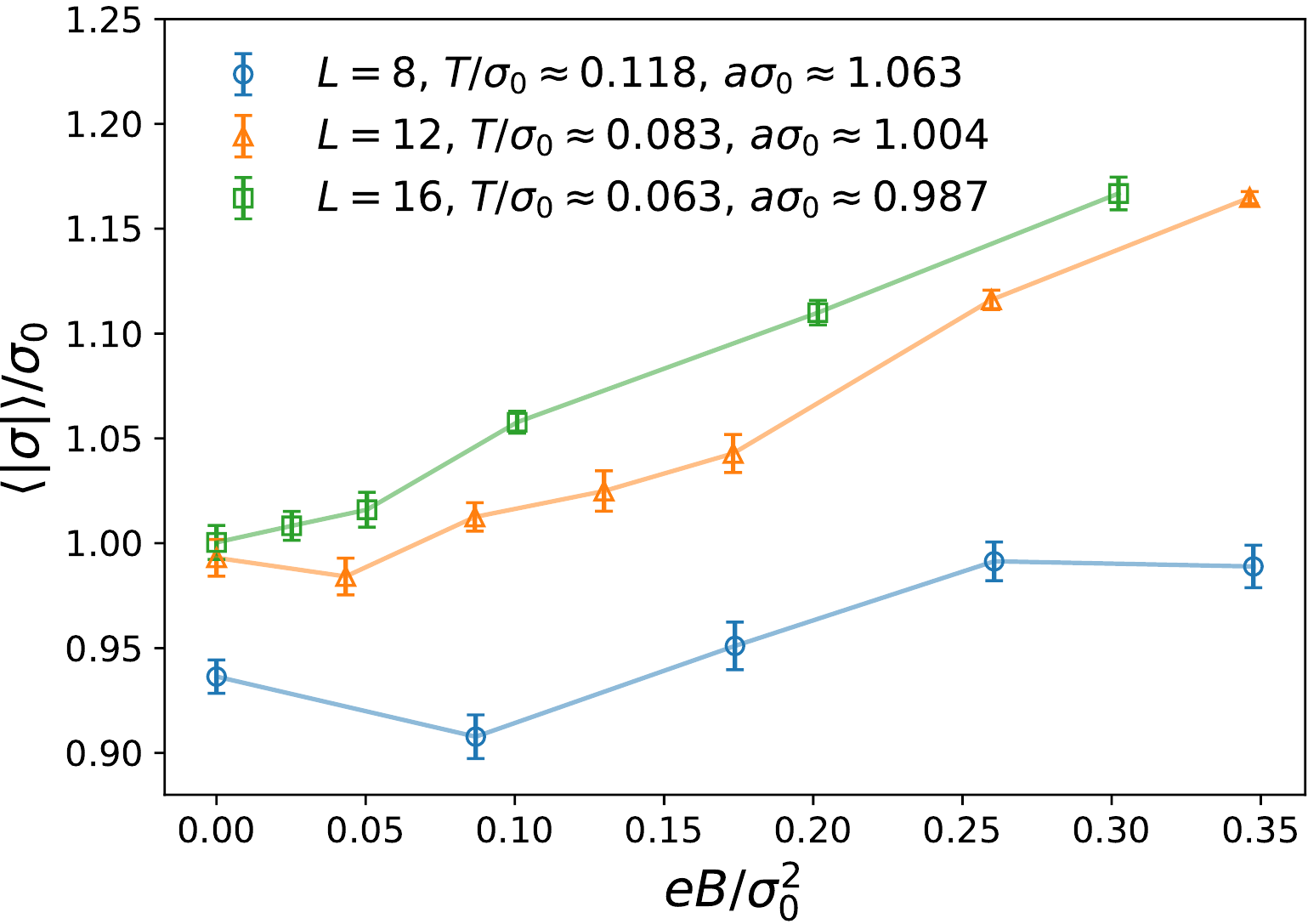}
		\caption{Infinite-volume extrapolation.}
		\label{fig:cc_vs_B_infVol}
	\end{subfigure}
	\caption{$B$-dependence of the order parameter at low temperatures and vanishing chemical potential.}
		\label{fig:cc_vs_B}
\end{figure}

Having gained some confidence in our discretization, we now consider a finite external magnetic field and study the $B$-dependence of the chiral condensate for low temperatures and vanishing chemical potential. We show both a continuum and an infinite-volume extrapolation in Figs.~\ref{fig:cc_vs_B_cont} and \ref{fig:cc_vs_B_infVol}, respectively. 

First looking at the former, it appears as if the condensate would decrease for the lowest non-vanishing value of the magnetic field (corresponding to $b=1$ in Eq.~\eqref{eq:magnetic_field}), before increasing again for all larger values of $B$. We can, however, by taking into account larger volumes in Fig.~\ref{fig:cc_vs_B_infVol}, clearly identify this as a finite-size effect\footnote{A more detailed analysis on this issue will be presented in an upcoming publication.} as it gradually disappears as the volume is increased. On the largest volume we observe clear evidence for magnetic catalysis, i.e., a monotonic increase of $\langle\vert\sigma\vert\rangle$ with $B$. This result is in qualitative agreement with the behavior found in mean-field \cite{Kli92_2} as well as beyond-mean-field \cite{KPR13} calculations, which does not come as a surprise, as it has been argued to be a universal, model-independent feature \cite{GMS95}.

\begin{figure}[h]
	\centering
	\includegraphics[scale=0.6]{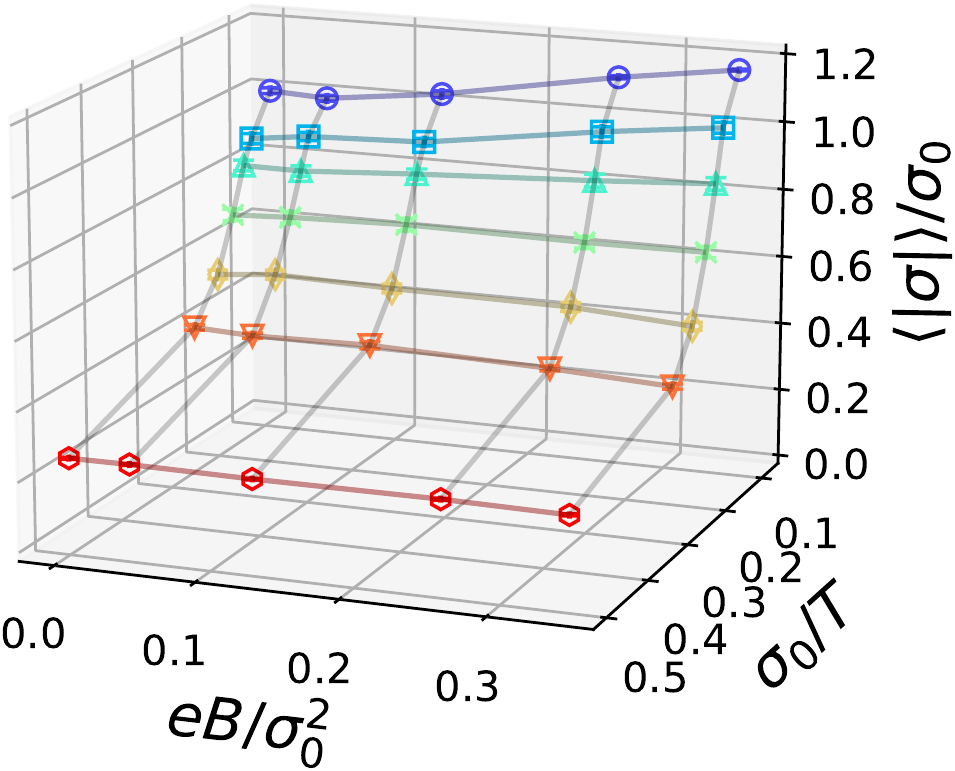}
	\caption{$T$-dependence of the order parameter as a function of the magnetic field for $\Ns=12$, $\mu=0$ and $a\sigma_0\approx 1.004$.}
		\label{fig:catalysis_T}
\end{figure}

The situation may, in principle, change for higher temperatures. In fact, in QCD one observes a decrease of the chiral condensate with the magnetic field around the transition temperature \cite{BBE12_2}. This inverse magnetic catalysis, however, is caused by a delicate interplay between quarks and gluonic degrees of freedom \cite{BEK13}. Hence, in accordance with mean-field results \cite{Kli92_2}, we should not expect to observe this effect in our setup, due to the absence of gluons. We confirm this claim by investigating how the $T$-dependence of the condensate changes with the magnetic field in Fig.~\ref{fig:catalysis_T}.

More precisely, we observe that for low temperatures, i.e., within the broken phase, the magnetic field causes an increase of the chiral condensate. Close to and beyond the phase transition, however, the condensate is largely independent of $B$. This is consistent with the expectation since magnetic catalysis, obviously, only occurs in the broken region. The latter should, in principle, expand with $B$, but this effect is likely negligible for the small magnetic fields we consider. We have checked this by studying the $B$-dependence of the critical temperature $T_c$ of the chiral phase transition, determined via the peak of the chiral susceptibility
\begin{align}
	\chi = \langle\sigma^2\rangle-\langle\vert\sigma\vert\rangle^2
\end{align}
as a function of $T$. We find that, indeed, the value of the critical temperature, $T_c\approx0.14\sigma_0$, changes only very little with $B$.

By performing the appropriate extrapolations, we find that the qualitative features of the chiral condensate at finite $T$ and $B$ shown in Fig.~\ref{fig:catalysis_T} persist when going to the infinite-volume and continuum limits, with the finite-size effects mentioned above again disappearing for large volumes. 

\section{Summary \& Outlook}
In this contribution we have investigated the phase diagram in the $(T,B)$ plane of the $(2+1)$-dimensional Gross-Neveu model \eqref{eq:4F_Lagrangian} beyond the mean-field limit for one reducible fermion flavor on the lattice, using overlap fermions. We find that the magnetic catalysis phenomenon, i.e., an enhancement of chiral symmetry breaking due to an applied magnetic field, predicted by large-$\Nf$ studies, persists when taking into account bosonic quantum fluctuations at $\Nf=1$. This is consistent with analytical calculations going beyond the mean-field approximation using optimized perturbation theory (OPT) to investigate the $\Nf=2$ case \cite{KPR13}. 

Clearly, this result does not represent QCD, where gluonic degrees of freedom play a significant role in the phase structure, giving rise to the inverse catalysis phenomenon at finite temperature. Going forward, there are many ways to systematically improve the degree to which we can approximate QCD using toy models. For instance, one could attempt to incorporate interactions with gluons by coupling the fermions to the Polyakov loop. Combined with the introduction of a suitable magnetic-field-dependence of the coupling constant, this has been shown to reproduce known QCD features in the mean-field Nambu--Jona-Lasinio (NJL) model in \cite{EM19}. The NJL model is, in a sense, an extension of the GN model, exhibiting a continuous chiral symmetry as well as three more bosonic degrees of freedom, arising from an additional pseudoscalar-isovector interaction channel. In another step towards QCD these pseudoscalar degrees of freedom could, together with the $\sigma$ field, furthermore be endowed with kinetic terms to allow for their interpretation as dynamical mesons.

Since ultimately we would like to study QCD at finite density we will perform a similar analysis to the one presented in this work also at finite chemical potential. When studying the system in the $(\mu, B)$ plane, one, however, has to deal with a complex-action problem and a more extensive analysis is required. Our aim will be to investigate whether the inverse magnetic catalysis and multiple phase transitions that occur in the mean-field limit persist when bosonic fluctuations are taken into account. That this should indeed be the case has been argued in \cite{KPR13}.

\acknowledgments

We wish to thank Laurin Pannullo, Marc Wagner and Marc Winstel for sharing their insights on four-Fermi theories on various occasions and Marc Winstel for valuable comments on this manuscript. M. M. thanks Gergely Endr\H{o}di and Tam\'as Kov\'acs for useful discussions. The authors are indebted to Björn Wellegehausen for providing the simulation code base used in this work. This work has been funded by the Deutsche Forschungsgemeinschaft (DFG) under Grant No. 406116891 within the Research Training Group RTG 2522/1. The numerical simulations were performed on resources of the Friedrich Schiller University of Jena supported in part by the DFG grants INST 275/334-1 FUGG and INST 275/363-1 FUGG. 

\bibliographystyle{JHEP}
\bibliography{bibliography}

\end{document}